\documentclass[12pt]{article}
\usepackage{epsfig}

\begin{document}

\title{Solutions of the  dispersion equation in the region of
overlapping of zero-sound and particle-hole modes}
\author{V. A. Sadovnikova,\\
Petersburg Nuclear Physics Institute, \\
Gatchina, St.~Petersburg 188300,
Russia}
\date{}
\maketitle
\begin{abstract}
In this paper the solutions of the zero-sound dispersion equation in the random phase approximation (RPA) are considered.
The calculation of the damped zero-sound modes $\omega_s(k)$ (complex frequency of excitation) in the nuclear matter is presented. The method is based on the analytical structure of the polarization operators $\Pi(\omega,k)$.\\
The solutions of two dispersion equations with $\Pi(\omega,k)$ and with ${\rm Re}\,(\Pi(\omega,k))$ are compared. It is shown that in the first case we obtain one-valued smooth solutions without "thumb-like" forms. \\
Considering the giant resonances in the nuclei as zero-sound excitations we  compare the experimental energy and  escape width of the giant dipole  resonance (GDR) in the nucleus  A  with $\omega_s(k)$ taken at a definite wave vector $k=k_A$.
\end{abstract}

\section{Introduction}
The investigation of the zero-sound dispersion equation was started in the Landau kinetic theory \cite{LaLi,PN}. One of the direction of the  further development of this theory consisted in application of RPA to the problem of the zero-sound mode propagation \cite{PN,FW}. The problem of the propagation and spreading of the zero-sounded  excitations in the various media, at the  different temperatures and energies,  is now a subject of the active investigations  \cite{4}-\cite{8}.

In this paper we consider the  problem of zero-sound excitations in the framework of the standard RPA \cite{LaLi,PN,
FW}. We reproduce the well-known results for the excitations with the  real frequencies (for the small wave vectors $k$). The goal of this note is to continue calculations to the larger $k$ when the  zero-sound $\omega_s(k)$ and the quasiparticle-quasihole $1p$ -$1h$ modes overlap. In the region of overlapping we have  the complex zero-sound frequencies $\omega_s(k)$.

The polarization operator $\Pi(\omega,k)$ has two cuts, which are related to the appearance of the real $ph$-pairs. The overlapping of $\omega_s(k)$ and $ph$-pairs means that the zero-sound solution (its real part ${\rm Re}\,\omega_s(k)$) is placed on the cut of the polarization operator. The imaginary part ${\rm Im}\,\omega_s(k)$ shows the shift of the solution from the real axis. There are two possibilities: 1)~$\omega_s(k)$ is shifted to the lower part of the physical sheet below the cut and 2)~$\omega_s(k)$ is shifted to the unphysical sheet neighboring with the physical one through the cut. We will try to demonstrate that the second possibility is realized.

The imaginary part ${\rm Im}\,\omega_s(k)$ is determined by  the processes which result in the appearance of the cut in the polarization operator. It means that  the physical reason of appearance  of ${\rm Im}\,\omega_s(k)$ is the emerging of the real $ph$-pairs when the real particle leaves the nucleus. Therefore we can interpret this ${\rm Im}\,\omega_s(k)$ as  the escape width of the zero-sound excitations.

We compare the solutions of two dispersion equations, one with the full polarization operator $\Pi(\omega,k)$ and another  with  the real part of polarization operator ${\rm Re}\,(\Pi(\omega,k))$. It is shown that there is an important distinction between the solutions in the region with the nonzero ${\rm Im}\,(\Pi(\omega,k))$.  The dispersion equation with the full $\Pi(\omega,k)$ has the smooth solutions $\omega_s(k)$ which can be continued far enough (till the application of the RPA is valid ). The imaginary part of $\omega_s(k)$ emerges  when the overlapping of the solutions and the cut of polarization operator takes place and is a smooth slowly decreasing function. In contrary, solutions of the dispersion  equation with  ${\rm Re}\,(\Pi(\omega,k))$ have the "thumb-like" form and disappear  at a special value of the wave vector, at this point the imaginary part
go to the infinity.

From the very beginning of the application of the Landau Fermi liquid theory to the nuclei \cite{Mi} it was shown that the giant resonances in nuclei can be considered as zero-sound excitations \cite{MZL,SW}. There are a lot of calculations of the energy and the width of the collective states and the giant resonances in the nuclei in the framework of the theory of the finite fermi systems \cite{Mi}. The relation of RPA to the Landau Fermi liquid theory in the  nuclei is demonstrated in \cite{Sp}.
 The theoretical consideration of the giant resonances can be characterized by some approaches.
The first one considers the  nuclei with  the nuclear mean field potentials and different forms of the residual interaction.   It uses the wave functions and the energies of the one-particle states with the further diagonalization of the RPA matrix \cite{SW,IYE}. The second approach is related to the  description of nuclei with the collective variables in the framework of the quantum hydrodynamic model \cite{SW,IYE}.

In this paper
we use the Fermi liquid theory \cite{Mi,Sp} to obtain  the zero-sound dispersion equation in the nuclear matter. Further the application of this equation solutions to the concrete nuclei is made. The branch of solutions of the zero-sound dispersion equation in the nuclear matter, $\omega_s(k)$, permits us to evaluate the energies and widths of the giant resonances in nuclei. Using the Steinwedel-Jensen model \cite{RS,BV} we can find a value of the wave vector $k=k_A$ corresponding to the giant resonances in the nucleus A. In the next step, $\omega_s(k_A)$ can be compared with the experimental giant resonance energies. In this work the comparison is made for the giant dipole resonances. The imaginary part of $\omega_s(k)$ is the result of the zero-sound decay to the real particle-hole pairs and can be compared with the escape width of resonances.

In  Sect.~2 the explanation of our approach is given. In  Sect.~3  the dispersion equation solutions with the full polarization operator $\Pi(\omega,k)$ and with ${\rm Re}\,(\Pi(\omega,k))$ are compared.
In Sect.~4  we evaluate  the energies and widths of the giant dipole resonances using  $\omega_s(k)$ and compare the result with the known formula fitting the experimental GDR energies.

\section{Method of calculation}
We present a method of calculation of the zero-sound  solutions $\omega_s(k)$ of the Eq.~(\ref{1}), which are real or complex-valued depending on a value of the wave vector $k$.
We use the standard RPA  to obtain the retarded polarization operators $\Pi(\omega,k)$ \cite{LaLi,FW}.
The zero-sound dispersion equation in the symmetric nuclear matter is
\begin{equation}\label{1}
1 =  C_0 F\Pi(\omega,k).
\end{equation}
The factor $C_0=N_0^{-1}$, where $N_0=2p_Fm/\pi^2$, is the density of  states for  two sorts of the nucleons on the Fermi surface;
$F$ is the any of the Landau-Migdal dimensionless parameters of the quasiparticle-quasihole interaction ${\cal F}$ \cite{Mi}:
\begin{equation}\label{17}
{\cal F}(\vec\sigma_1,\vec \tau_1;\vec\sigma_2,\vec \tau_2) = C_0\left(F_0 + F'(\vec\tau_1\vec\tau_2) + G(\vec\sigma_1\vec\sigma_2)
 + G' (\vec\tau_1\vec\tau_2)\, (\vec\sigma_1\vec\sigma_2)\right),
\end{equation}
where $\vec\sigma$, $\vec \tau$ are the Pauli matrices in the spin and isospin spaces.
The polarization operator $\Pi(\omega,k)$ is the integral over a nucleon particle-hole loop:
\begin{equation}\label{9a}
\Pi(\omega,k) =
4 \int \frac{d^3p}{(2\pi)^3}\left[ \frac{\theta(|\vec p+\vec
k|-p_F)\theta(p_F-p)} {\omega+\frac{p^2}{2m}-\frac{(\vec p+\vec
k)^2}{2m}+i\eta} - \frac{\theta(p-p_F) \theta(p_F-|\vec p+\vec
k|)} {\omega+\frac{p^2}{2m}-\frac{(\vec p+\vec k)^2}{2m}+i\eta}
\right]
= -4\Phi(\omega,k).
\end{equation}
The factor 4 comes from the summation over the spin and isospin of the loop. The integral is taken by the traditional way \cite{FW}.
The function $\Phi(\omega,k)$ can be represented in the form
\begin{equation}\label{9}
\Phi(\omega,k) =
\phi(\omega,k) + \phi(-\omega,k).
\end{equation}

The function $\phi(\omega,k)$ is the Migdal's function \cite{Mi}
\begin{equation} \label{11}
\phi(\omega,k) =\ \frac1{4\pi^2}\
\frac{m^3}{k^3}
\left[\frac{a^2-b^2}2 \ln\left(\frac{a+b}{a-b}\right)-ab\right] ,
\end{equation}
where $a=\omega-(k^2/2m)$, $b=kp_F/m$ \footnote{This presentation of $\phi(\omega,k)$ differs from the one which was used in \cite{VS}, but the results for the branches of solutions do not depend on the chosen form of $\phi(\omega,k)$ in (\ref{9},\ref{11}). In \cite{VS} we used the form which is more convenient in the study of the relation of the long wavelength instability to the pion condensation.}.

Recall that the functions $\phi(\omega,k)$  and $\phi(-\omega,k)$ (\ref{11}) have the following  cuts on the complex $\omega$-plane.
The cut of $\phi(\omega,k)$ is placed at
\begin{equation} \label{12}
-\frac{kp_F}{m} + \frac{k^2}{2m}\ \le\ \omega\ \le\ \frac{kp_F}{m}+\frac{k^2}{2m}\ .
\end{equation}
The cut of $\phi(-\omega,k)$ is placed at
\begin{equation} \label{12p}
-\frac{kp_F}{m} - \frac{k^2}{2m}\ \le\  \omega\ \le\ \frac{kp_F}{m} - \frac{k^2}{2m}\ .
\end{equation}
These overlapping cuts of the retarded polarization operator $\Pi(\omega,k)$ are shown in Fig.1.
The polarization operator $\Pi(\omega,k)$ has the nonzero imaginary part, ${\rm Im}\,(\Pi(\omega,k))$, for the frequencies $\omega$ in the intervals (\ref{12}), (\ref{12p}).
On the one hand,  ${\rm Im}\,(\Pi(\omega,k))$ appears as an imaginary part of the logarithm on the cut in (\ref{12}).
On the other hand, the same ${\rm Im}(\Pi(\omega,k))$ can be interpreted as a result of the excitation of the real $ph$-pairs for the given $\omega$, $k$, $p_F$, when the function $-i\pi\delta(\omega+\frac{p^2}{2m}-\frac{(\vec p+\vec k)^2}{2m})$  contributes to integral (\ref{9a}).

The physical sheet is defined by the condition that in the long wavelength limit $k\to 0$ and $\frac{\omega}{k}\to const$, we obtain the physical sheet of the kinetic theory of Landau.  As a result the imaginary part of the logarithm in Eq.(11) in the complex $\omega$-plane is equal to $(-i\pi)$ on the upper edge of the cut Eq.(\ref{12}) and $(+i\pi)$ on the lower edge  \cite{PN}.
When the wave vector increases, the cuts overlapping becomes shorter and disappears at $k=2p_F$. At $k>2p_F$ there are two nonoverlapping cuts in $\Pi(\omega,k)$.

For not too large wave vectors $k$ the collective excitations have the  frequencies which are larger than the $ph$-pair ones. It means that $\omega_s(k)$ lies on the real axis on the complex $\omega$-plane more to the right than the cut (\ref{12}). With increasing $k$, the cut overlaps with the solutions $\omega_s(k)$. The overlapping takes place at a specific wave vector  $k_{th}$ when the value of $\omega_s(k_{th})$ touches the cut Eq.(\ref{12}) at $\omega= \frac{k_{th}p_F}{m}+\frac{k_{th}^2}{2m}$. Substituting $\omega_s(k_{th})$ in Eq.(\ref{1}) as solution we obtain
\begin{equation}\label{3}
-\frac{2+F}{F} = (x+1)\ln\left(\frac{x}{x+1}\right),
\end{equation}
where $x=\frac{k_{th}}{2p_F}$. For $F=2$ we obtain $k_{th}=0.51~p_0$ ($p_0$=268~MeV).

How can we continue our solution at $k>k_{th}$?  At these wave vectors the dispersion equation (\ref{1}) acquires the large imaginary part and we expect that the solutions are the imaginary ones.   We are interested in the solutions with the negative imaginary part ${\rm Im}\,\omega_s(k)$. There are two possibilities, the first: the branch $\omega_s(k)$ stays on the physical sheet and goes under the lower edge of cut and the second: $\omega_s(k)$ goes through the cut to the unphysical sheet.

 The logarithm $\ln\left( \frac
{\omega m+kp_F+\frac12k^2}{\omega m-kp_F+\frac12k^2}\right)\equiv \ln(Z)$ has the value $\ln(|Z|)-i\pi$ on the upper edge of the cut (\ref{12}) and the value
$\ln(|Z|)+i\pi$ on the lower edge. Thus, on the physical sheet the discontinuity of logarithm over the cut is equal to $(+2\pi i)$. Under the cut there is an  unphysical sheet neighboring with the physical one on the Riemann surface of $\ln(Z)$. The magnitude of $\ln(Z)$ on the neighboring sheet differs by  $(-2\pi i)$. Thus, $\pi i$ is  canceled and we have a continuous change of logarithm, when  go  under the cut (following the arrow in Fig.~1). Then one may conclude that $\Pi(\omega,k)$ in Eq.~(\ref{1}) changes continuously as well.

To determine what of the possibilities realizes, we rewrite the Eq.~(\ref{1}) in the following way
\begin{equation}\label{13}
\left(\left(\frac{1}{C_0F}\frac{1}{(-4)}-\phi(-\omega,k)\right) \times \left[\frac1{4\pi^2}\
\frac{m^3}{k^3}\right]^{-1} + ab\right) \times \left[\frac{a^2-b^2}2\right]^{-1} =  \ln\left(\frac{a+b}{a-b}\right).
\end{equation}
 We look for the solutions of Eq.~(\ref{1}) and (\ref{13}) with the negative imaginary part: $\omega_s= \omega_r +i\omega_i$ and $\omega_i<0$.
Let calculate the left (LHS) and right (RHS) sides of Eq.(\ref{13}) at a definite $k$ and $\omega_i$ changing $\omega_r$. In Fig.~2 the real and the imaginary parts of LHS is shown by the solid curves (numbers '1' and '2'). The dashed curves are presented for the real and imaginary parts of RHS (curves '3' and '4'), the dotted curve is $\rm Im(RHS)$ on the nearest unphysical sheet (curve '5'). We see that the imaginary part of LHS (curve '2') is negative. The $\rm Im(RHS)$ (curve '4') is positive on the physical sheet under the cut. It becomes negative on the unphysical sheet (curve '5'). The solution is the crossing of the real parts of LHS (curve '1') and RHS (curve '3') and their imaginary parts (curve '2' and curve '4' or '5') at the same value of $\omega_r$. In Fig.~2 we see the crossing of curves '2' and '5' therefore the logarithm must be taken on the unphysical sheet. The detailed discussion of Fig.~2 is done in Appendix.

It is of importance that moving on the sheets of the
Riemann surface of the one of the logarithms, we remain on the  physical sheet of the other logarithm.
The branch $\omega_s(k)$ goes under the cut (\ref{12}) (Fig.~1) and not under  the symmetric cut (\ref{12p}). We add a term $(-2\pi i)$ to the logarithm $\ln(Z)$  but not to another logarithm.
To calculate $\omega_s(k)$ on the unphysical sheet, we solve Eq.~(\ref{1}) with the shifted logarithm.
If we are interested in the symmetric branch $\omega'_s(k)$ [${\rm Re}\,(\omega'_s(k))= -{\rm Re}\,(\omega_s(k))$ and  ${\rm Im}\,(\omega'_s(k))$= ${\rm Im}\,(\omega_s(k))$], then at  $k>k_{th}$ it goes through the symmetric cut (\ref{12p}).

The branch $\omega_s(k)$ is presented in Fig.~3 by the solid curves. We see that $\omega_s(k)$ is real at the wave vectors $k\leq k_{th}$ and the imaginary part of $\omega_s(k)$ (shown in the bottom) appears at $k > k_{th}$.
The dashed curves denote the cuts. The space between the $x$-axis and the lower dashed curve is the region where $\Pi(\omega,k)$ gets the imaginary part due to the overlapping cuts (\ref{12}), (\ref{12p}).  The space between two  dashed curves is the region where $\Pi(\omega,k)$ gets the imaginary part due to the cut (\ref{12}).

In this paper  a simple model with the constant effective quasiparticle-quasihole interaction (\ref{17}) is used. In Fig.~3 the solutions of Eq.~(\ref{1}) are presented for the symmetric nuclear matter  at the equilibrium density, corresponding to the Fermi momentum $p_F$=$p_0$=268~MeV.  We take the quasiparticle mass $m= 0.8m_0$ ($m_0=940$~MeV) \cite{Mi,Sp}. The parameter $F$ is a constant and  $F$=2. This value of $F$ corresponds to the isospin $F'$ and spin-isospin $G'$ parameters  in the the Landau-Migdal effective interaction (\ref{17}) (recalculated to the value  of $C_0$ of this paper).

\subsection{Long wavelength limit}
In the long wavelength  limit the dispersion equation
(\ref{1}) with Eqs.~(\ref{9}), (\ref{11}) turns to the well-known Landau dispersion equation.
For $F>0$ we can find $k_{th}$ in (\ref{3}). For $k<k_{th}$ the solutions are real and  close to the solutions of Landau dispersion equation.

\section{Dispersion equation with the real part of polarization operator}
It is instructive to compare $\omega_s(k)$ with the solutions of the widely used  dispersion equation including the real part of the polarization operator only \cite{7,8,AM} :
\begin{equation}\label{1r}
1 = C_0\, F\, \mbox{Re}(\Pi(\omega,k)).
\end{equation}
The dotted curve in  Fig.~3, having a "thumb-like" form, is the solution of this equation,  $\omega_{sr}(k)$.

We see that the upper part of the dotted curve and the solid one  are  identical when the polarization operator is real. But at $k>k_{th}$ there is a difference between  them.
The mode $\omega_s(k)$ becomes complex-valued. The real part  ${\rm Re}\,(\omega_s(k))$
continues to increase. The imaginary part  ${\rm Im}\,(\omega_s(k))$ is negative and smoothly decreases.
The solutions of Eq.(\ref{1r}) demonstrate  another dependence:
 $\omega_{sr}(k)$  decreases at $k>k_{th}$ and disappears at $k=k_{max}= 0.74~p_0$.
The attempts to find  additional solution  of the dispersion equation (\ref{1}) (i.e., with the full polarization operator) corresponding to the lower part of the dotted curve $\omega_{sr}(k)$ (Fig.~3) on the physical or neighboring unphysical sheets failed.

When the damping is small, we can use the following equation for the imaginary part of the zero-sound mode \cite{FW}:
\begin{equation}\label{14}
\omega_{si}(k) = \mbox{Im }\Pi(\omega_{sr},k)\left(\frac
{\partial \mbox{Re}\Pi)}{\partial\omega}\right)^{-1}_{\omega=\omega_{sr}(k)} .
\end{equation}
Here $\omega_{sr}(k)$ are  solutions of Eq.~(\ref{1r}), which are placed on the upper part of the "thumb-like" form in Fig.~3. In  Fig.~3, $\omega_{si}(k)$ is shown by the dotted curve.
We see that $\omega_{si}(k)$ and ${\rm Im}\,(\omega_s(k))$ coincide at a very small damping only.

In some papers  in which  the pion dispersion equations are considered \cite{8,AM}, the equations with the real part of the polarization operator (similar to Eq.(\ref{1r})) are used. The solutions have a complicated form with a lot of the "thumb-like" structures. The method  suggested in this paper, permits us to obtain several one-valued smooth  complex functions $\omega_{n}(k)$ instead of the many-valued "thumb-like" structures. Moreover the branches of the solutions $\omega_{n}(k)$ can be related to the solutions on the physical sheet by the uninterrupted continuation on $k$ or $p_F$ or $F$ \cite{VS}. Thus we know the physical nature of the every branch of solutions. Since the imaginary part of the solutions appears due to the passing to the unphysical sheet through the cut, the nature of the cut determines what kind of the width of excitations we calculate. This statement is valid, for example, for the zero-sound excitations in the nuclear matter in the spin-isospin channel when both the particle-hole and the isobar-hole cuts  exist.

\section{Giant resonances}
In this Section  the giant dipole resonances  are considered. In the simple version of the Migdal  particle-hole effective interactions  the electrical dipole collective states are excited by the  isovector interaction constant $F'$ (\ref{17}).
The coefficient $C_0$ is defined in (\ref{1}). The  values of $F'$ are changed in the region  from 0.8 to 1.9 \cite{Mi,Sp}  at $C_0=\pi^2/(2p_0m_0)$=150 MeV~fm$^3$.

First of all we obtain the solution of Eq.(\ref{1}) with $F=F'$.
The Steinwedel-Jensen model \cite{RS,BV} determines the wave vector $k_A$ which corresponds to the definite  excitations in the nucleus. We follow to \cite{BV}, where $k_A=\frac{\pi}{2R}$, where $R=r_0 A^{1/3}$, $r_0=1.2$~fm. Using this relation of $k$ to $A^{1/3}$ we get  $\omega_s(A)$. We see that $k$ is larger for the lighter nuclei (this is the consequence of the boundary conditions).

In the next step we compare  $\omega_s(A)$  with the formula for the GDR energies in the different nuclei \cite{Atomic}:
\begin{equation}\label{15}
E_{GDR}(A) = 31.2A^{-1/3} + 20.6 A^{-1/6} \mbox{MeV}.
\end{equation}

Before doing the comparison we investigate the dependence of the branch $\omega_s(k)$ on the nuclear matter parameters: $p_F$, $F$, $m$ which are included in Eq.~(\ref{1}) .
In Fig.~4 we consider the dependence of $\omega_s(k)$ on the constants of the effective quasiparticle interactions $F$ (Fig.~4a), on the value of Fermi momentum $p_F$ (in Fig.~4b) and on the effective mass $m$ (Fig.~4c).

In Fig.~4a we present the branches $\omega_s(k)$ for that values of $F$ which are obtained from the experiments \cite{Sp}.
In \cite{FW} it is demonstrated that the frequency of zero-sound excitations increases with increasing of $F$ (Fig.~16.1 \cite{FW}) at a fixed $k$.
We obtain the same behavior if to fix rather small $k$ and take for $F$ such values that  $\omega_s(k)$ is real. But in the region of overlapping of the zero-sound and particle-hole modes, when ${\rm Im}\,\omega_s(k)$ is nonzero, the behavior is changed, this is shown in Fig.~4a.
The growth of $F$ results in decreasing of both ${\rm Re}\,\omega_s(k)$ and ${\rm Im}\,\omega_s(k)$. The sensitivity of the real and imaginary parts of  $\omega_s(k)$ to the value of $F$ is different:  ${\rm Re}\,\omega_s(k)$ is almost equal  at $F=1$ and $F=2$ but ${\rm Im}\,\omega_s(k)$ is changed considerably.

In Fig.~4b we present the branches $\omega_s(k)$ for some values of $p_F$. The growth of $p_F$ from 200~MeV to 300~MeV results in the increasing of ${\rm Re}\,\omega_s(k)$ and decreasing of ${\rm Im}\,\omega_s(k)$.
In Fig.~4c we see that the growth of the effective quasiparticle mass from $m=600$~MeV to $m=940$~MeV give the decreasing of both ${\rm Re}\,\omega_s(k)$ and ${\rm Im}\,\omega_s(k)$.

Thus, we obtain a different dependence of the zero-sound  excitations on the nuclear matter parameters $F'$, $p_F$ and $m$. Changing these parameters we try to approximate  the phenomenological equation (\ref{15}) by $\omega_s(k_A)$.

In Fig.~5 the comparison of the experimental and calculated energies of the giant dipole resonances in the nuclei is presented. It is difficult to expect that  such a complex phenomenon as  giant resonances can be described  quantitatively by the simple model used in this paper. Nevertheless, we  expect that the main features are reproduced.
In Fig.~5 we present the curve  (\ref{15}) by the solid line. The $\rm{Re}\,\omega_s(k_A)$ correspond to the energies of GDR and are shown at the positive $\omega$. The $\rm{Im}\,\omega_s(k_A)$ which correspond to the half-width of the GDR are negative and are shown at the negative $\omega$.
The dashed-dot curves show the results for the same nuclear matter parameters as in Fig.~3 ($p_F=268$~MeV, $F$=2.0, $m=0.8m_0$). We see that the calculated  energies of resonances are too large and the width too small.

Now we try to find that parameters of the nuclear matter which permit us to describe, at least generally, the  GDR in the  nuclei.
To approximate  the phenomenological curve (\ref{15}) by our results we use two sets of the nuclear matter parameters. One set  is the  more suitable for the light nuclei and the second is better for the heavy ones.
The branch  $\omega_s(k_A)$ obtained with $p_F=200$~MeV, $F$=1.2, $m=m_0$ is shown by the dashed line.  By the dotted line we draw the $\omega_s(k_A)$ obtained with the  matter parameters  $p_F=200$~MeV, $F$=1.6, $m=0.8m_0$.

For the  nucleus $^{12}C$ we have from Eq.~(\ref{15}) $E_{GDR}=27.2$~MeV, the experimental escape width is $\Gamma\approx 3$~MeV.
On the dashed curve we obtain $E_{GDR}=30$~MeV and $\Gamma=6.8$~MeV. On the dotted curve: $E_{GDR}=37$~MeV and $\Gamma= 7.8$~MeV. For the  nucleus $^{208}Pb$ we have from Eq.~(\ref{15})  $E_{GDR}=13.7$~MeV, the experimental escape width is $\Gamma\approx 0.5-2.0$~MeV. On the dashed curve we obtain $E_{GDR}=10$~MeV and $\Gamma\approx 0.5$~MeV. On the dotted curve $E_{GDR}=12.7$~MeV and $\Gamma=0.5$~MeV. We obtain the better approximation to the curve Eq.~(\ref{15}) when start from the nuclear matter with rather low density:  $p_F$=200~MeV, $\rho/\rho_0\approx 0.42$. This points to the important contribution of the nuclear surface  to the excitation of the resonances.

In Fig.~5  the decreasing of the GDR energies and escape width with the growth of the atomic number A is demonstrated. In Fig.~3 we see the expected figure when the  $\rm{Re}\,\omega_s(k)$ and  $(-\rm{Im}\,\omega_s(k))$ increase with the growth of the wave vector $k$.
The Steinwedel-Jensen model \cite{RS,BV}  shows that the excitations with the same quantum numbers have the larger wave vectors $k$  in the lighter nuclei.  Thus  using the simple model of the zero-sound in the nuclear matter we obtain the qualitative description of the experimental fact that the energies and the escape width of the giant resonances in the light nuclei are larger than in the heavy ones.

\section{Discussion}
In this paper we demonstrate the method of calculation of the  zero-sound  frequencies (real or complex) in the region of overlapping  of zero-sound and particle-hole modes.
It is shown that the damped zero-sound solutions are placed on the unphysical sheet which is neighboring with the physical one on the Riemann surface of the logarithm in Eq.~(\ref{11}). The logarithmic cut (\ref{12}) corresponds to the excitation of   $ph$-pairs, and we can conclude that ${\rm Im}\,(\omega_s(k))$  is the half-width of  the zero-sound mode decay into  $ph$-pairs.

It is shown in the Fig.~3,  that  neglecting of the imaginary part of the polarization operator in the dispersion equation, we obtain  the branch of solutions having a "thumb-like" form with the upper point of the real part  and the quick decrease of the imaginary part at $k=k_{max}$ . On the other hand, considering the full polarization operator at a proper unphysical sheet, we have an continuous  behavior of the zero-sound mode $\omega_s(k)$ without  the singularities.

The zero sound branches $\omega_s(k)$ are calculated in the nuclear matter with the different values of the Fermi momentum $p_F$, the nucleon effective mass $m$ and the effective residual interaction $F$. The giant resonances can be considered as the zero sound excitations in the nuclei. Using the simple model of the quasiparticle interaction (\ref{17}) we obtain at least the qualitative description of the experimental fact that  the energies and the escape width of the giant resonances in the light nuclei are larger than in the heavy ones.

The zero sound mode $\omega_s(k)$  calculated in the nuclear matter in equilibrium ($p_F=268$~MeV, $F$=2.0, $m=0.8m_0$, $C_0$=150~MeV\,fm$^3$ in Eq.(\ref{1}))  corresponds to the very high energies and  low escape widths of GDR (dot-dashed curve in Fig.~5). The GDR in the nuclei correspond rather to the zero sound in the nuclear matter with lower density (the dashed and dotted curves in Fig.~5). This fact is probably related to the importance of the nuclear surface in forming of the resonances.

\section{Appendix}
In this Appendix we try to show that the solutions of the zero-sound dispersion equation are placed on the nearest unphysical sheet when the overlapping of the zero-sound and the $ph$ excitations take place.

In Fig.~2 the left and right sides of Eq.~(\ref{13}) are shown as  functions on the frequency. The calculation is made for the fixed wave vector $k$=0.65$p_0$. As it is shown in Fig.~3,  the solution of Eq.(\ref{1}) (and of (\ref{13}), identical to it) at $k$=0.65$p_0$ is equal to  $\omega_s(k=0.65p_0)=(0.304,-0.003)p_0$.
We choose  $k$=0.65$p_0$ which is close to $k_{th}$ in Eq.~(\ref{3}), and we see in Fig.~3 that the solution of (\ref{1}) is close to the solution of the approximate equations (\ref{1r}), (\ref{14}).

In Fig.~2 the imaginary part of $\omega$ is taken to be equal to the imaginary part of the solution $\omega_i$ =-0.003$p_0$. The solid curve '1' ('2') is the real (imaginary) part of LHS of Eq.~(\ref{13}). We see that for the negative $\omega_i$ the imaginary part of LHS is negative also.  The dashed curve '3' is the real part of logarithm in the RHS of (\ref{13}) and the curve '4' is its imaginary part on the physical sheet. As it was  discussed above the imaginary part of logarithm on the physical sheet on the lower edge of the cut is positive and close to $+i\pi$. The dashed curve '5' is the imaginary part of logarithm on the nearest unphysical sheet of the Riemann surface, where the imaginary part is differ by $-2i\pi$ and is negative. In Fig.~2 a letter  'A' ('B') denotes the crossing of the real (imaginary)  parts of LHS and RHS of (\ref{13}) at $\omega_r=0.304p_0$, that corresponds to the solution of (\ref{1}). In Fig.~2 we see  that  the imaginary parts of LHS  crosses  that branch of logarithm in RHS which is place on the unphysical sheet.


\newpage

\section{Figure captions}
\noindent
FIG.~1. Complex plane of the frequency  $\omega$.
The cut (\ref{12})  is shown from 1 to 1', the cut (\ref{12p}) is shown from 2 to 2'.

\noindent
FIG.~2.  Comparison of LHS and RHS of Eq.~(\ref{13}). Solid curves: '1' presents $\rm Re$\,(LHS) and '2' shows $\rm Im$\,(LHS).
Dashed curves: '3' is $\rm Re$\,(RHS), '4' is $\rm Im$\,(RHS) on the physical sheet, '5' is $\rm Im$\,(RHS) on the unphysical sheet. The figure is made at $k=0.65p_0$ and $\omega_i=-0.003p_0$. A letter  'A' ('B') denotes the crossing of the real (imaginary)  parts of LHS and RHS  at $\omega_r=0.304p_0$. $p_0$=0.268~MeV.

\noindent
FIG.~3. Zero-sound branch of solutions $\omega_s(k)$ of Eq.~(\ref{1}) (solid curves). The $\omega_{sr}(k)$ are the solutions of Eq.~(\ref{1r}) and $\omega_{si}(k)$ are solutions of (\ref{14}) (dotted curves).
Dashed lines are the boundary of the cuts  (\ref{12}), (\ref{12p}); $k_{th}$ denotes the wave vector when the overlapping of $\omega_s(k)$ and quasiparticle continuum starts.

\noindent
FIG.~4. Zero-sound branch of solutions $\omega_s(k)$ for the different values of the effective constant $F$ (figure $a$), the Fermi momentum $p_F$ (figure $b$) and the effective mass $m$ (figure $c$).
${\rm Re}\,(\omega_s)$ are drawn at the positive $\omega$, ${\rm Im}\,(\omega_s)$ are drawn at the negative $\omega$. The blobs mark $k_{th}/p_0$  for the curves 1,2,3.
Fig.~$a$: $F$=0.2, 1.0, 2.0 (numbers 1,2,3, correspondingly).
Fig.~$b$: $p_F$=200, 268, 300~MeV (numbers 1,2,3, correspondingly).
Fig.~$c$: $m$=500, 750, 940~MeV (numbers 1,2,3, correspondingly).

\noindent
FIG.~5. Comparison of the experimental GDR energies with the calculated ones ($A$ is the atomic number). At the positive $\omega$ we compare $E_{GDR}$ (\ref{15}) with $\rm{Re}\,\omega_s(A)$. At the negative $\omega$ the $\rm{Im}\,\omega_s(A)$ are presented.
The solid curve shows the approximation (\ref{15}). Dashed curves stand for $\omega_s(A)$ calculated for $p_F$=200~MeV, $m=m_0$ and $F$=0.6. Dotted curves are calculated  with $p_F$=200~MeV, $m=0.8m_0$ and $F$=0.5.
Dash-dotted curves are for $p_F$=268~MeV, $m=0.8m_0$ and $F$=2.


\begin{figure}
\centering{\epsfig{figure=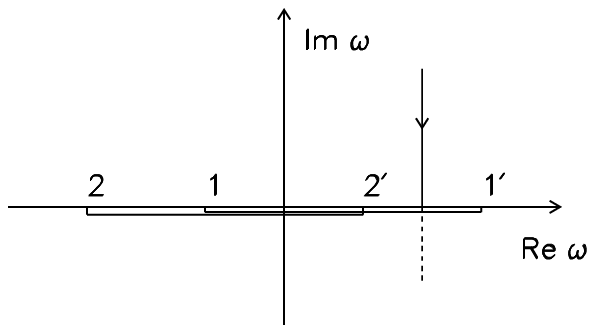,width=7cm}}
\caption{}
\end{figure}

\begin{figure}
\centering{\epsfig{figure=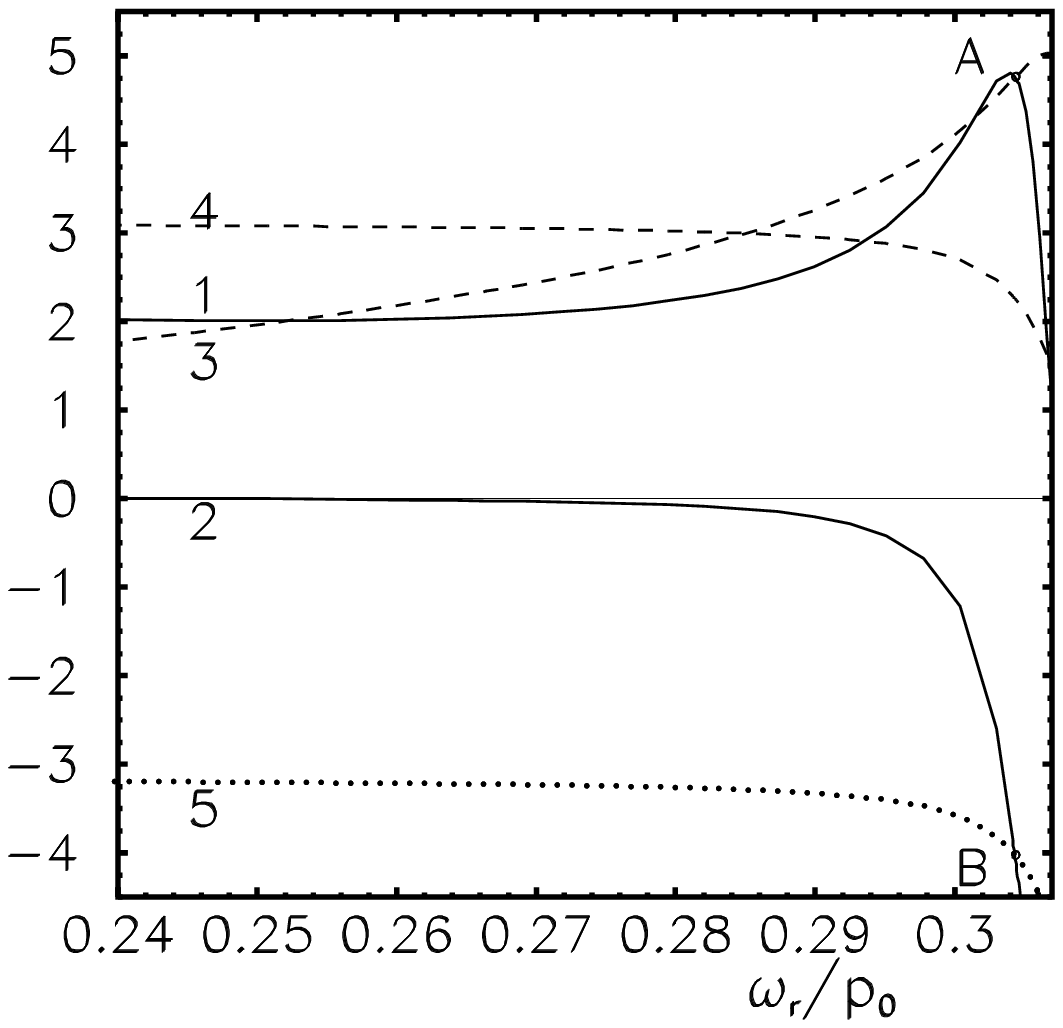,width=7cm}}
\caption{}
\end{figure}

\begin{figure}
\centering{\epsfig{figure=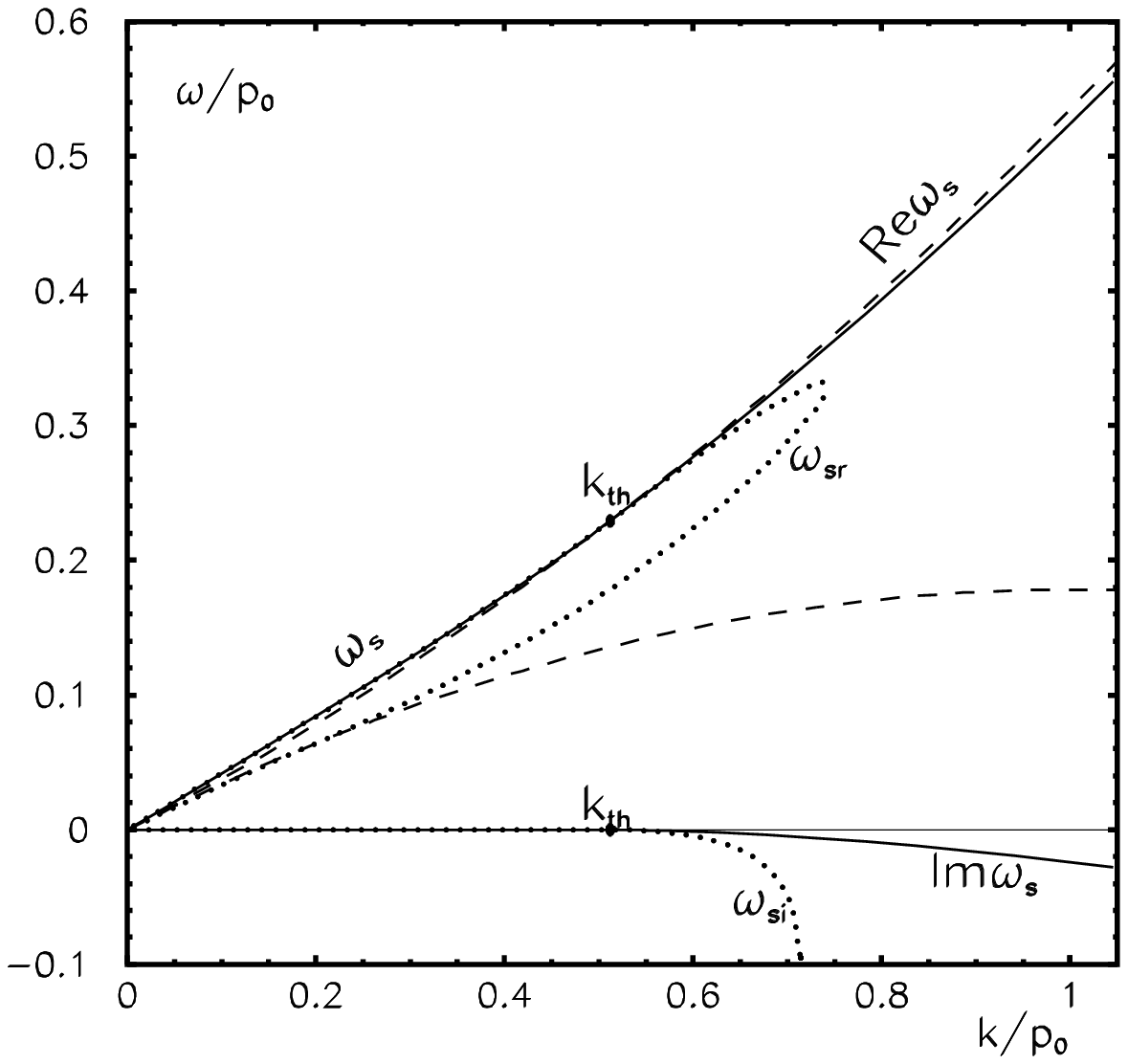,width=9cm}}
\caption{}
\end{figure}


\begin{figure}

\vspace{1.0cm}

\hspace{3.5cm}  $a$ \hspace{6.5cm}  $b$

\vspace{-1.0cm}

\centering{\epsfig{figure=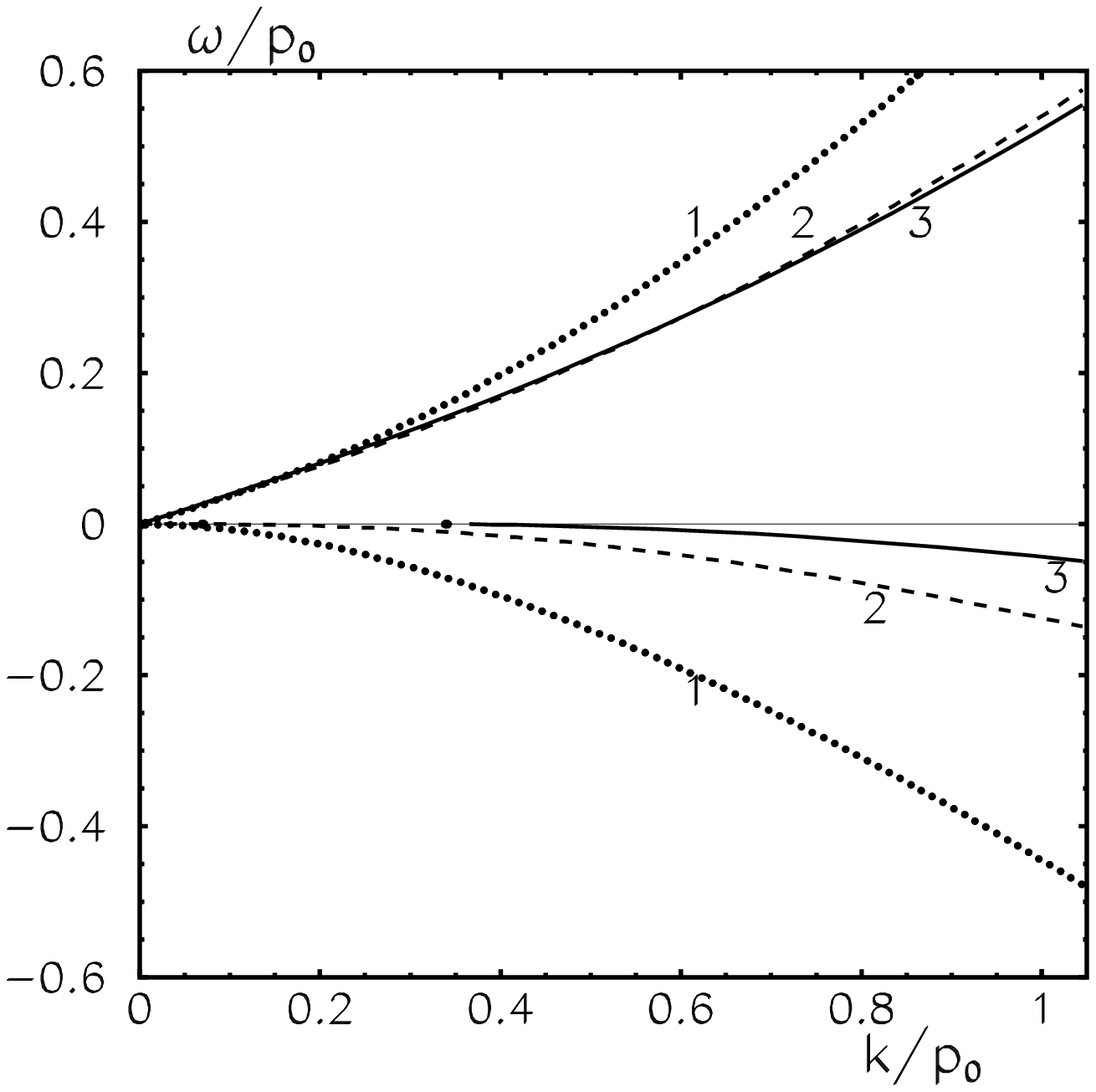,width=7cm}\epsfig{figure=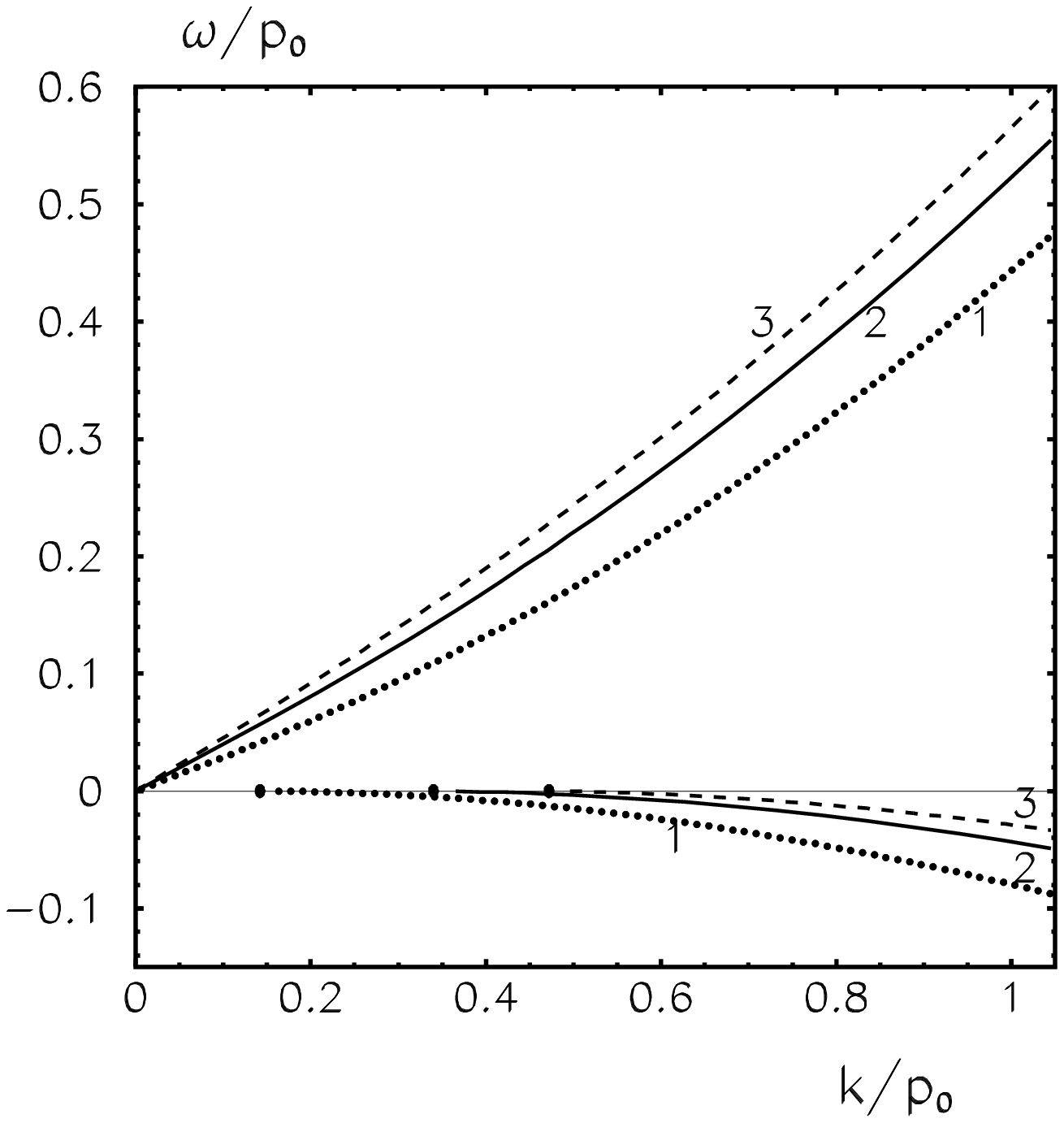,width=7cm}}

 \vspace{1cm}

\hspace{-1cm}  $c$

\vspace{-1cm}

\centering{\epsfig{figure=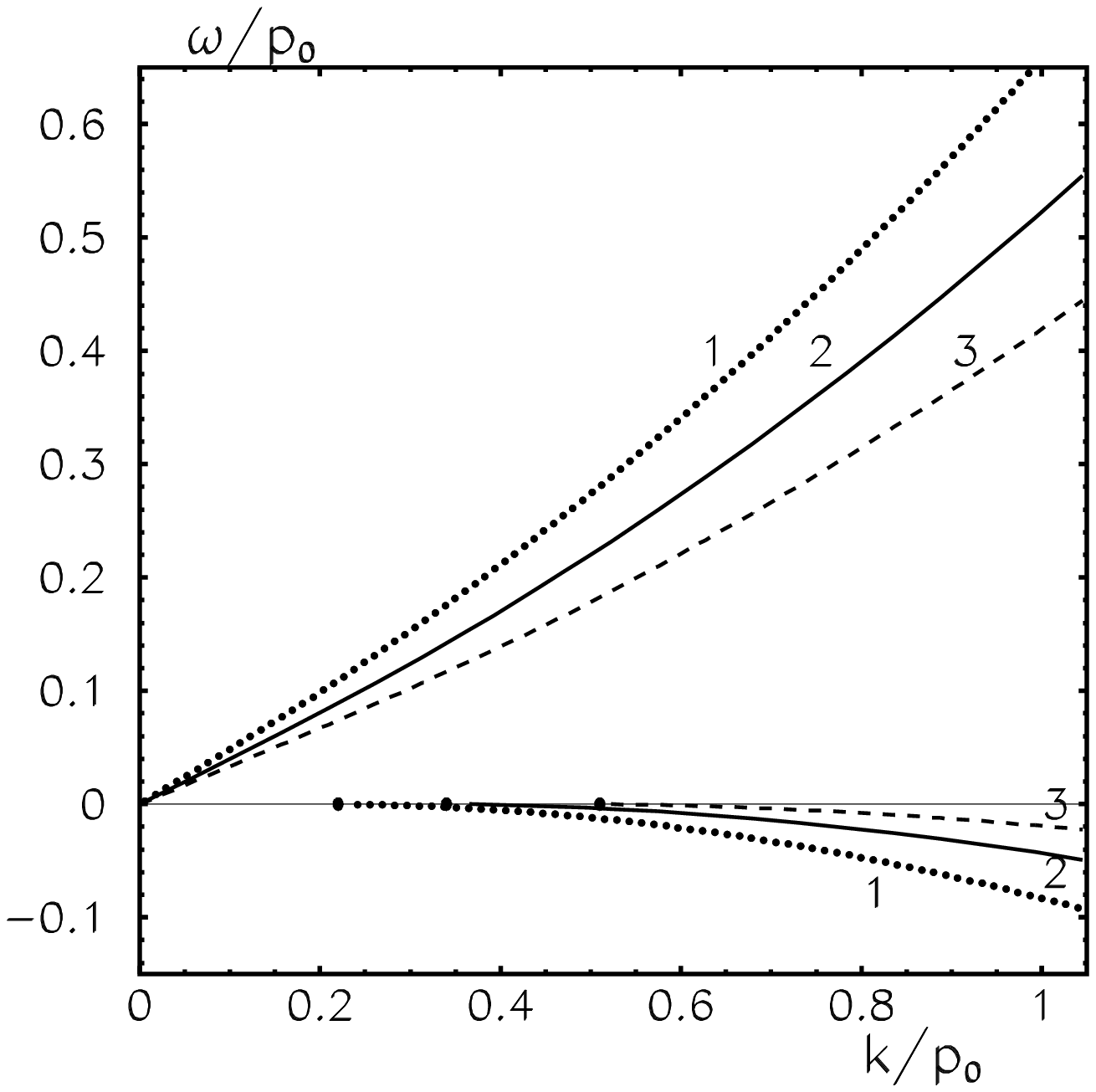,width=7cm}}

\caption{}
\end{figure}

\begin{figure}
\centering{\epsfig{figure=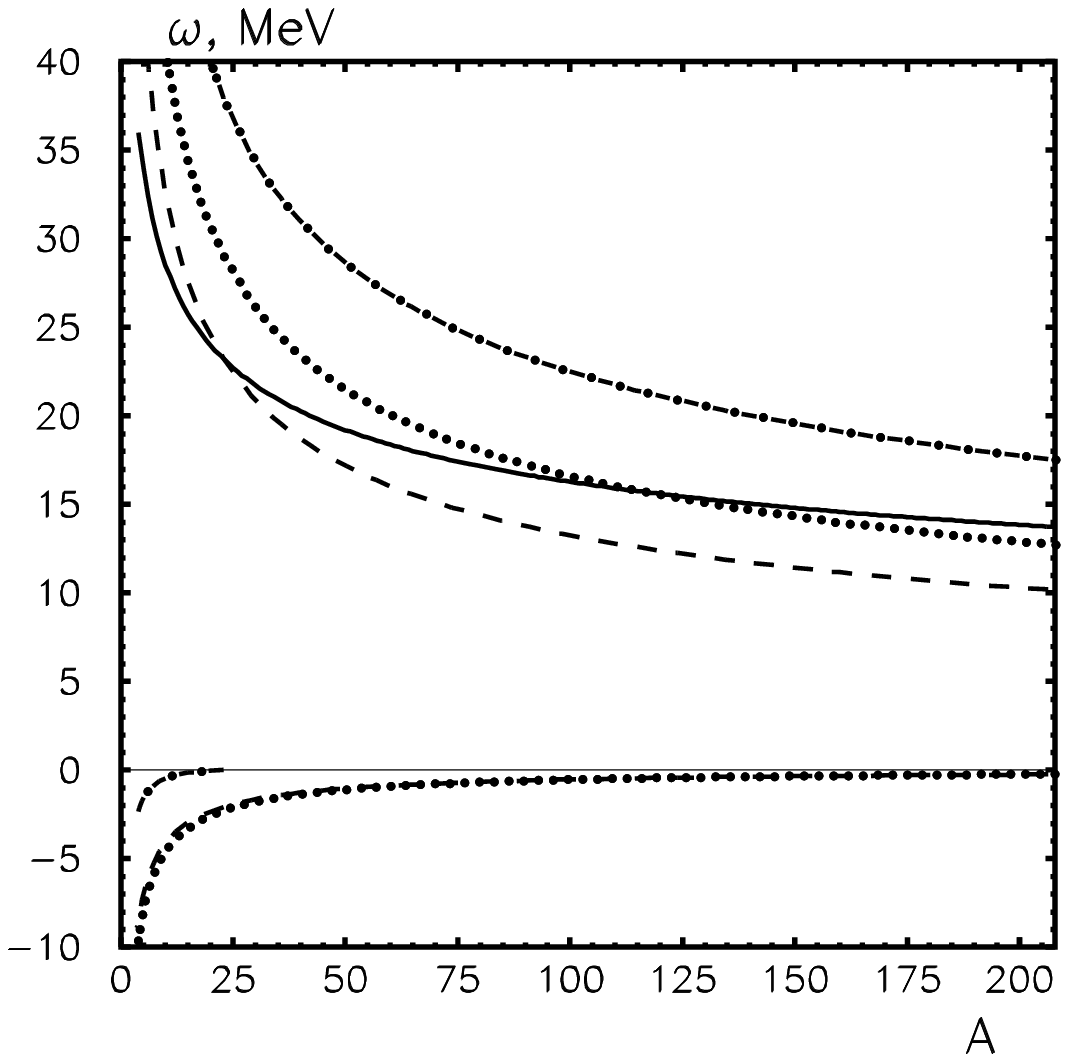,width=8cm}}
\caption{}
\end{figure}


\newpage

\begin{thebibliography}{**}

\bibitem{LaLi} L. D. Landau and E. M. Lifshitz,  {\em``Statistical
physics"} (Pergamon Press, 1989).

\bibitem{PN} D. Pines and P. Nozieres,
{\em``The Theory of Quantum Liquids"} (W.A.Benjamin,Inc., 1966).

\bibitem{FW} A. L. Fetter and J. D. Walecka,{\em``Quantum theory of
many-particle systems"}(Mc-Graw-Hill, New York, 1971).


\bibitem{4} C. Toepffer, P. Hiltner, Z. Phys. {\bf B74}, 429 (1989).

\bibitem{5} A. B. Larionov, J. Piperova, M. Colonna, M. Di Toro, Phys.  Rev. {\bf C61}, 064614  (2000).

\bibitem{6} V. M. Kolomietz, S. Shlomo, Phys. Rev. {\bf C64},  044304 (2001).

\bibitem{7} M. Baldo, C. Ducoin,  Phys. Rev. {\bf C79}, 035801 (2009).

\bibitem{8} L.-G. Liu, M. Nakano, Nucl. Phys. {\bf A618}, 337 (1997).

\bibitem{Mi} A. B. Migdal, {\em``Theory of Finite  fermi Systems and Properties of the Atomic Nucleus"} (Willey, New-York,1967; 2 Ed. Nauka, Moskow, 1983); Rev. Mod. Phys., {\bf 50}, 107 (1078);
  A.~B.~Migdal, D.~N.~Voskresenskii, E.~E.~Saperstein, and
M.~A.~Troitskii, Phys. Rep. {\bf 192}, 179 (1990).

\bibitem{MZL} A.B. Migdal, D.F. Zaretsky, A.A. Lushnikov, Nucl. Phys., {\bf 66}, 193 (1965).

\bibitem{SW} J. Speth, J. Wambach in book {\em``Electric and magnetic giant resonances in nuclei"},
Ed. by J. Speth, 1991, World Sientific Publishing Company.

\bibitem{Sp} J. Speth, E. Werner, W. Wild, Phys.Rep. {\bf 33}, 127 (1977);
 S. O. Backman,  G. E. Brown,  J. A. Niskanen, Phys. Rep. {\bf 124}, 1 (1985).


\bibitem{IYE} B.S. Ishkhanov, N. P. Yudin, R. A. Eramzhyan, PEPAN, {\bf 31}, 313 (2000).

\bibitem{RS} P. Ring, P. Schuck, {\em``The nuclear many-body Problem"},
 (Springer-Verlag, 1980).

\bibitem{BV} F.~L. Braghin, D. Vautherin, Phys. Lett. {\bf B333}, 189 (1994).

\bibitem{Atomic} Atomic Data and Nuclear Data Tables, {\bf 15}, 319 (1975).


\bibitem{VS} V.~A.~Sadovnikova, Phys. Atom. Nucl., {\bf 70}, 989 (2007).

\bibitem{AM} J. Diaz Alonso and A.~Perez~Canyellas, Nucl. Phys.
{\bf A526}, 623 (1991).

\end{thebibliography}
\end{document}